\documentclass[aip,
 amsmath,amssymb,
reprint,%
]{revtex4-2}

\usepackage{graphicx}
\usepackage{dcolumn}
\usepackage{bm}

\usepackage[utf8]{inputenc}
\usepackage[T1]{fontenc}
\usepackage{mathptmx}
\usepackage{etoolbox}

\usepackage{physics} 
\usepackage{amssymb} 

\newcommand{\xloc}{\mathbf{x}}
\newcommand{\xs}{\mathbf{x}_{s}}
\newcommand{\xss}{\mathbf{x}_{ss}}

\newcommand{\xt}{\mathbf{x}_{t}}
\newcommand{\hhat}{\hat{h}}

\newcommand{\loperator}{\mathcal{L}}
\newcommand{\mubar}{\bar{\mu}}
\newcommand{\force}{\mathbf{f}}

\newcommand{\identity}{\mathbf{I}}

\newcommand{\localop}{\boldsymbol{\Lambda}}
\newcommand{\Uvelocity}{\mathbf{U}_{0}}
\newcommand{\hhatsim}{\hhat_{sim}}

\usepackage{color}

\makeatletter
\def\@email#1#2{%
 \endgroup
 \patchcmd{\titleblock@produce}
  {\frontmatter@RRAPformat}
  {\frontmatter@RRAPformat{\produce@RRAP{*#1\href{mailto:#2}{#2}}}\frontmatter@RRAPformat}
  {}{}
}%
\makeatother
\begin{document}

\title{Flow-Induced Buckling of Elastic Microfilaments with Non-Uniform Bending Stiffness}
\author{Thomas Nguyen} 
\author{Harishankar Manikantan}%
\affiliation{ 
Department of Chemical Engineering, UC Davis, Davis, CA
}%
 \email{hmanikantan@ucdavis.edu}

\date{\today}

\begin{abstract}
\vspace{6mm}
Buckling plays a critical role in the transport and dynamics of elastic microfilaments in Stokesian fluids. However, previous work has only considered filaments with homogeneous structural properties. Filament backbone stiffness can be non-uniform in many biological systems like microtubules, where the association and disassociation of proteins can lead to spatial and temporal changes into structure. The consequences of such non-uniformities in the configurational stability and transport of these fibers are yet unknown.  Here, we use slender-body theory and Euler-Bernoulli elasticity coupled with various non-uniform bending rigidity profiles to quantify this buckling instability using linear stability analysis and Brownian simulations. In shear flows, we observe more pronounced buckling in areas of reduced rigidity in our simulations. These areas of marked deformations give rise to differences in the particle extra stress, indicating a nontrivial rheological response due to the presence of these filaments. The fundamental mode shapes arising from each rigidity profile are consistent with the predictions from our linear stability analysis. Collectively, these results suggest that non-uniform bending rigidity can drastically alter fluid-structure interactions in physiologically relevant settings, providing a foundation to elucidate the complex interplay between hydrodynamics and the structural properties of biopolymers.

\vspace{10mm}

\end{abstract}

\maketitle


\section{\label{sec:introduction}Introduction}


Elastic filaments such as actin and microtubules serve as the backbone of cells by providing structural integrity to the intracellular matrix. Beyond this core function, fluid-structure interactions between these elastic fibers and their surrounding fluid are essential to many biological processes. One such example is cytoplasmic streaming, a process where motor protein movement in filament networks can drive fluid flow within cells.\cite{Verchot_Lubicz2010,Suzuki2017,Woodhouse2013} Advances in microfluidic methods and the rich nonlinear dynamics that arise from the fluid-structure interactions have spurred many experimental, analytical and numerical investigations into the elastohydrodynamics of single filament systems.\cite{Kantsler2012,Wiggins1998,Chelakkot2010,Chelakkot2012,Harasim2013,Chakrabarti2020,Chakrabarti2020_RCS}

Like the deformation of Euler beams, elastic filaments moving freely in viscous fluids can undergo a buckling instability if the compressive forces acting upon the filament exceed the internal restorative elastic forces. This phenomenon has been well characterized in cellular, \cite{Young2007,Wandersman2010,Quennouz2015} extensional, \cite{Kantsler2012,Manikantan2015} shear, \cite{Becker2001, Liu2018} and other flow profiles. \cite{Chakrabarti2020,Chakrabarti2020_RCS} Thermal fluctuations due to Brownian motion add additional subtleties to this elastohydrodynamic problem: experiments have demonstrated the rounding of this buckling instability,\cite{Kantsler2012} which has been analytically substantiated as well. \cite{Baczynski2008,Manikantan2015} This well-characterized instability is crucial in dictating the transport of these filaments. Bending and buckling allows fibers to move as random walkers near hyperbolic stagnation points in a 2D cellular flow array,  \cite{Young2007} whereas thermal fluctuations hinder the transport of these fibers and trap them within the vortical cells.\cite{Manikantan2013} More recently, filaments have been shown to be trapped around circular objects due to flow-induced buckling. \cite{Chakrabarti2020_RCS} 

However, in all of these studies, the bending stiffness or rigidity of the filament backbone is assumed to be uniform. This may not hold true in many physical systems. For example, protein adsorption onto filaments can be highly non-uniform, following heterogeneous condensation that has been linked to the Rayleigh-Plateau instability. \cite{Hernandez_Vega2017,Setru2021} This non-uniformity has been characterized in recent experiments that correlate regions of increased microtubular bending with enhanced protein adsorption. \cite{Tan2019} Some past studies have modeled filaments with heterogeneous mechanical properties and predicted their resulting deformation and fragmentation behavior, \cite{DeLaCruz2015,Lorenzo2020} but it is unclear what shapes and configurations these filaments can assume. A platform to predict the expected filament shapes and buckling thresholds of non-uniformly stiff filaments in fluid flow has not yet been established. Moreoever, a quantitative analysis of the growth of buckling modes is still missing. In this work, we present results from linear stability analysis and non-linear simulations for heterogeneously stiff filaments undergoing the buckling instability in flow.

In what follows, we provide a mathematical description of filaments with non-uniform and heterogeneous rigidity profiles coupled to slender-body theory for viscous flows. In addition to a constant bending rigidity traditionally used in conjunction with slender-body theory, we analyze two examples of asymmetrical rigidity profiles with analytical forms motivated by protein attachment/detachment onto/from filaments. We also lay out the process to determine fundamental modes or shapes for any rigidity profile, and provide a consistent framework to extract amplitudes of these modes from future simulations and experiments. We use this platform to report qualitative and quantitative differences in linear stability analyses and nonlinear Brownian simulations across select non-uniform stiffness profiles.

\section{\label{sec:problem_desc}Problem Description}

Describing the dynamic conformations of flexible filaments in flow requires solving the Navier-Stokes equation coupled to elastic equations for the filament backbone. The length and time scales are such that the Navier-Stokes equation reduces to the Stokes equation: classic works on slender-body theory for low-Reynolds-number hydrodynamics then describe the fluid dynamics of the problem if the force distribution corresponding to the presence of a filament are known. \cite{Batchelor1970,Keller1976,Johnson1980} For this work, we will use the local Brownian motion variation of slender-body theory to couple microfilament dynamics with a viscous fluid.\cite{Tornberg2004,Manikantan2013, Manikantan2015} This variation accounts for the anisotropy of the filament based on its shape and orientation, but neglects non-local hydrodynamic interactions between different points on the filament. 

\subsection{\label{sec:energy_func} Energy Functional and Force Balance} 


We will consider an inextensible elastic filament of characteristic thickness $2a$ and length $L$ which is parameterized by arclength $s$. Here, $s$ is a material parameter for the filament that serves to discretize the filament backbone from $-L/2$ to $L/2$ and thus is independent of time $t$. The slenderness of the filament is captured by the slenderness ratio $\epsilon= a/L \ll 1$. The centerline coordinates of the filament are $\xloc(s,t) = (x(s,t),y(s,t),z(s,t))$; for the purposes of this work, we shall consider flow and buckling in the 2D $x$--$y$ plane. When a non-Brownian filament is placed in flow, the competition between the external viscous forces and internal elastic or tensile forces describes its dynamics. We follow the approach used by Li \textit{et al.}\ \cite{Li2013} to derive a non-uniform force across such a filament, starting with the energy functional:
\begin{equation}\label{eq:filament_elastic_energy}
\mathcal{E} = \frac{1}{2} \displaystyle \int^{L}_{0}{(\kappa(s) \xss^{2} + T(s)\left( \xs \cdot \xs - 1 \right)) \: ds} - 
\displaystyle \int^{L}_{0}{\force(s) \cdot \xloc(s) \: ds}.
\end{equation}
Here, $\kappa (s)$ is the non-uniform stiffness or bending modulus of the filament ($\kappa (s) = E(s)I$ where $E (s)$ is Young's Modulus and $I = \pi a^{4}/4$ is the second moment of inertia of a rod), $\xss$ is the filament curvature, $T$ represents the line tension experienced by the filament which enters as a Lagrange multiplier, and $f$ is the force per unit length exerted on the filament by the fluid. All subscripts on variables represent partial derivatives with respect to the subscript unless otherwise stated.

Physically, the terms in the first integral of Equation~(\ref{eq:filament_elastic_energy}) corresponds to an energetic penalty for bending and stretching, respectively. The last integral term relates the energy to the force at a particular location on the filament. To obtain this force, we can take variational derivatives of the energy functional above via the Euler-Lagrange Equation:
\begin{equation}\label{eq:euler_lagrange}
\pdv{\mathcal{E}}{\xloc} - \pdv{}{s}
\left(\pdv{\mathcal{E}}{\xs}\right) + \pdv[2]{}{s}
\left(\pdv{\mathcal{E}}{\xss}\right) = 0.
\end{equation}
This gives the dimensional force acting on the filament in the absence of Brownian motion:
\begin{equation}\label{eq:non_constant_K_force_dimensional}
\force (s) =  -(T (s) \xs)_{s} + (\kappa (s) \xss)_{ss}.
\end{equation}

\subsection{\label{sec:sbt_motion_eq}Constitutive equations of motion}

The filament is immersed in a fluid of viscosity $\mu$ with an imposed velocity field $\Uvelocity (\xloc (s,t),t)$. The velocity of the filament is then approximated by the local version of the slender-body theory centerline equation:\cite{Batchelor1970,Keller1976,Johnson1980, Tornberg2004,Manikantan2013,Manikantan2015,Chakrabarti2020_RCS}
\begin{equation}\label{eq:filament_velocity_dimensional}
8 \pi \mu \left(\xt(s,t) - \Uvelocity\left(\xloc,t\right)\right) = -\localop [\force].
\end{equation}
Here, $\mathbf{x}_{t}$ is the time derivative of the filament centerline, $\localop$ is a local operator that captures the filament's anisotropic interaction with the surrounding fluid, and $\force$ is the force per unit length acting on the filament as given by Equation~(\ref{eq:non_constant_K_force_dimensional}). The local operator is given by
\begin{equation}\label{eq:local_op}
\localop[\force](s)  = \left((c+1)\textbf{I} + (c-3)\xs\xs\right) \cdot \force,
\end{equation}
where $\xs \xs$ is the dyadic product of the unit vectors $\xs (s)$ that are locally tangent to the filament centerline and $c=\ln(1/\epsilon^{2})$.

In the absence of Brownian motion, we non-dimensionalize Equations~(\ref{eq:non_constant_K_force_dimensional}) and~(\ref{eq:filament_velocity_dimensional}) with length scales over $L$, time with the fluid flow strength $\dot{\gamma}$, and forces with a characteristic elastic force $\kappa/L^{2}$. These quantities collect into a single dimensionless parameter:
\begin{equation}\label{eq:mu_bar_definition}
\mubar = \frac{8\pi\mu \dot{\gamma} L^{2}}{\kappa/L^{2}},
\end{equation}
which can be interpreted as the ratio of viscous forces to elastic forces. Non-dimensionalization of Equations~(\ref{eq:non_constant_K_force_dimensional}) and ~(\ref{eq:filament_velocity_dimensional}) respectively results in the following dimensionless centerline velocity and force expressions:
\begin{subequations}\label{eq:non_brownian_dimensionless_all}
\begin{equation}
\mubar \left(\xt(s,t) - \Uvelocity\left(\xloc (s),t\right)\right) = -\localop [\force],
\end{equation}\label{eq:filament_velocity_non_brownian_dimensionless}
\begin{equation}
\force(s) = -(T\xs)_{s} + \left(B (s) \xss\right)_{ss},
\end{equation}\label{eq:filament_force_non_brownian_dimensionless}
\end{subequations}
where $B (s)$ is now the dimensionless stiffness profile across the filament. All variables hereafter are assumed to be dimensionless unless otherwise stated.

\begin{figure}[b]
 \centering
 \includegraphics[height=3.25cm]{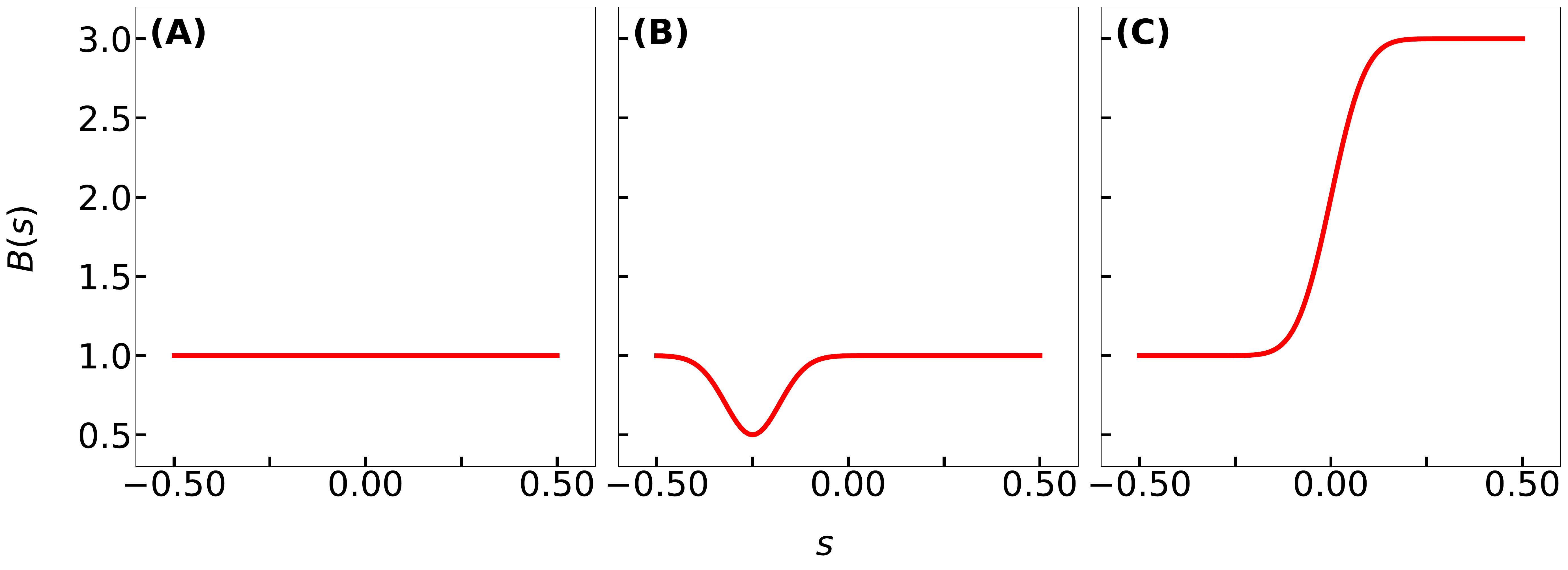}
 \caption{Rigidity profiles examined in this work. A) Constant rigidity $B_{1}(s)$. B) Locally weak stiffness profile $B_{2}(s)$ to reflect structural degradation due to protein attachment. C) Asymmetrically rigid stiffness profile $B_{3}(s)$ to reflect higher resistance to bending due to selective protein attachment one one half of the filament.}
 \label{fig:3_rigidity_profiles}
\end{figure}

\subsection{\label{sec:stiffness_profiles}Biologically motivated stiffness profiles}

Protein attachment onto microtubules and other elastic filaments can be highly non-uniform\cite{Hernandez_Vega2017,Tan2019,Setru2021}, resulting in heterogeneous structural properties. We are interested in the stability and configurations of filaments with non-uniform stiffness, which could shed light into their transport across streamlines and in complex flows. Motivated by protein adsorption or desorption that locally strengthens or weakens the filament backbone, we use the following analytical forms of bending stiffness profiles (Figure~\ref{fig:3_rigidity_profiles}):
\begin{subequations}\label{eq:stiffness_profile__all}
\begin{equation}\label{eq:stiffness_profile_constant}
B_{1}(s) = 1,
\end{equation}
\begin{equation}\label{eq:stiffness_profile_locally_weak}
B_{2}(s) = 1-\frac{1}{2}e^{-100\left(s+\frac{1}{4}\right)^{2}},
\end{equation}
\begin{equation}\label{eq:stiffness_profile_asym_rigid}
B_{3}(s) = 2+\erf(10s).
\end{equation}
\end{subequations}
The constant stiffness profile $B_{1}(s)$ is traditionally used by slender-body theory calculations, and we use this to test our predictions against previous works. \cite{Manikantan2015,Chakrabarti2020} The asymmetrical stiffness profiles $B_{2}(s)$ and $B_{3}(s)$ reflect potential protein adsorption patterns that locally modifies filament's stiffness. $B_{2}(s)$ was chosen to model the potential stiffness profile for a locally weak and asymmetric backbone that results in the the ``fish hook''-like microtubule in Tan \textit{et. al}'s study.\cite{Tan2019}. $B_{3}(s)$, on the other hand, may represent a situation where one half of a microtubule is weakened (or, equivalently, stiffened) due to protein condensation. Note, however, that the framework we develop below works for any fitted or modeled form of $B(s)$: the choices in Equations~\eqref{eq:stiffness_profile__all} are merely illustrative examples that we use to demonstrate our methods.

\subsection{\label{sec:brownian}Brownian Motion} 

Microscopic objects suspended in a fluid medium are subjected to Brownian forces: these thermal fluctuations are characterized by $k_{B}T$ where $k_{B}$ is Boltzmann's constant and $T$ is the absolute temperature. The elastic resistance of elongated structures to bending due to fluctuations is characterized by the persistence length $\ell_{p}=\kappa/k_{B}T$. Alternatively, $\ell_{p}$ a measure of distance between two points on an object at which the local tangent vectors become uncorrelated due to thermal fluctuations. Microtubules have a persistence length of approximately 5 mm, \cite{Gittes1993} which is roughly $\mathcal{O}(100-1000)$ times larger than their typical lengths. Actin filaments are more easily deformed by Brownian fluctuations, with $\ell_p/L=\mathcal{O}(1-10)$.\cite{Gittes1993} 

The stochastic Brownian force enters as an additional term in the dimensional force expression in Equation \eqref{eq:non_constant_K_force_dimensional}:
\begin{equation}\label{eq:non_constant_K_Brownian_dimensional}
\force (s) =  -(T (s) \xs)_{s} + (\kappa (s) \xss)_{ss} + \force^{Br}(s,t).
\end{equation}
We set up Brownian forces $\force^{Br}$ to satisfy the fluctuation-dissipation theorem of statistical mechanics that describes a fluctuating force with zero mean and finite variance proportional to $k_{B}T$ and the hydrodynamic resistance:
\begin{subequations}\label{eq:fluctuation_dissipation_all}
\begin{equation}\label{eq:fluctuation_dissipation_mean}
\langle \force^{Br}(s,t)\rangle = 0,
\end{equation}
\begin{equation}\label{eq:fluctuation_dissipation_variance}
\langle \force^{Br}(s,t)\force^{Br}(s',t')\rangle = 2k_{B}T \textbf{M}^{-1} \delta(s-s')\delta(t-t').
\end{equation}
\end{subequations}
Here, $\langle\rangle$ represents an ensemble average, $\delta$ is the dirac delta function, and $\mathbf{M}$ is the dimensional mobility tensor $(8\pi\mu)^{-1}((c+1)\identity + (c-3)\xs\xs)$ from Equation~(\ref{eq:local_op}). Non-dimensionalizing Equations~(\ref{eq:filament_velocity_dimensional}) and~(\ref{eq:non_constant_K_Brownian_dimensional}) now requires recognizing the large scale separation between the flow and Brownian time scales.\cite{Manikantan2013,Manikantan2015, Munk2006} To accommodate filament deflections at these Brownian time scales, we use the relaxation time of the elastic filament $8\pi\mu L^{4}/\kappa$ to nondimensionalize time scales associated with filament movement. The external flow field is still scaled over $L\dot{\gamma}$, and lengths and forces are nondimensionalized like before. The resulting dimensionless constitutive equations for Brownian motion are then:
\begin{subequations}\label{eq:brownian_dimensionless_all}
\begin{equation}\label{eq:filament_velocity_brownian_dimensionless}
\xt(s,t) - \mubar\Uvelocity\left(\xloc (s),t\right) = -\localop [\force],
\end{equation}
\begin{equation}\label{eq:force_brownian_dimensionless}
\force(s) = -(T\xs)_{s} + \left(B (s) \xss\right)_{ss} + \sqrt{\frac{L}{\ell_{p}}}\boldsymbol{\xi}(s).
\end{equation}
\end{subequations}
where $\boldsymbol{\xi}(s)$ is the dimensionless Brownian force. 
\subsection{\label{sec:numerical_methods}Numerical and Computational Methods}

We employ numerical methods that have been previously described in detail and tested across various flow geometries. \cite{Tornberg2004, Manikantan2013,Manikantan2015,Chakrabarti2020_RCS} Briefly, Equations~(\ref{eq:filament_velocity_non_brownian_dimensionless}) and~(\ref{eq:filament_velocity_brownian_dimensionless}) contain an unknown line tension that is first determined by applying the identity $(\xs \cdot \xs)_{t} = 0$. The resulting differential equation in $T(s)$ is solved using tension-free boundary conditions: $T|_{s=\pm 1/2} = 0$. Then, the filament position is solved with the torque-free and force-free boundary conditions:\cite{Landau1986}
\begin{subequations}\label{eq:torque_force_boundary_condition_all}
\begin{equation}\label{eq:torque_force_boundary_condition_lhs}
B(s)\xss|_{s=-\frac{1}{2}} = B(s)\xss|_{s=+\frac{1}{2}} = 0,
\end{equation}
\begin{equation}\label{eq:torque_force_boundary_condition_rhs}
\left(B(s)\xss\right)_{s}\mid_{s=-\frac{1}{2}} = \left(B(s)\xss \right)_{s}\mid_{s=+\frac{1}{2}} = 0.
\end{equation}
\end{subequations}
All spatial and time derivatives are approximated with second-order finite difference approximations.\cite{Tornberg2004} However, the fourth derivative term in the elastic term in the expression for the force exerted on the filament enforces a stringent restriction on the time step size. This problem is mitigated with a semi-implicit time marching scheme previously developed by Tornberg \& Shelley.\cite{Tornberg2004} The work presented here uses a slenderness ratio $\epsilon$ of 0.01. 

Brownian forces are calculated from Equation~(\ref{eq:force_brownian_dimensionless}) using previously established methods. \cite{Manikantan2013,manikantan_phd2015} We numerically evaluate $\boldsymbol{\xi}(s)$ as:
\begin{equation}\label{eq:dimensionless_Brownian_value}
\boldsymbol{\xi}(s) = \sqrt{\frac{2}{\Delta s \Delta t}}\: \textbf{B} \cdot \boldsymbol{\omega}.
\end{equation}
Here, $\boldsymbol{\omega}$ is a random vector from a Gaussian distribution of zero mean and unit variance, $\Delta s$ is the grid spacing of the filament, $\Delta t$ is the time step size, and $\textbf{B}$ is the tensor square root of $\textbf{M}^{-1}$ such that $\textbf{B}\boldsymbol{\cdot}\textbf{B}^T=\textbf{M}^{-1}$. We verify our results with the uniform stiffness profile against established past works involving non-Brownian \cite{Tornberg2004} as well as Brownian \cite{Manikantan2015} filaments in shear flow.

\section{\label{sec:results}Results}

We will first present our numerical results of non-linear simulations of a non-Brownian fiber. We then use linear stability analysis to help explain our numerical findings and establish key differences between the stiffness profiles. Following this, we compare our linear mode predictions from stability analyses with the filament shapes observed at the onset of instability in non-linear and Brownian simulations.

\subsubsection{\label{sec:simulation_results}Simulation Results}

\begin{figure}[t]
 \centering
 \includegraphics[height=12cm]{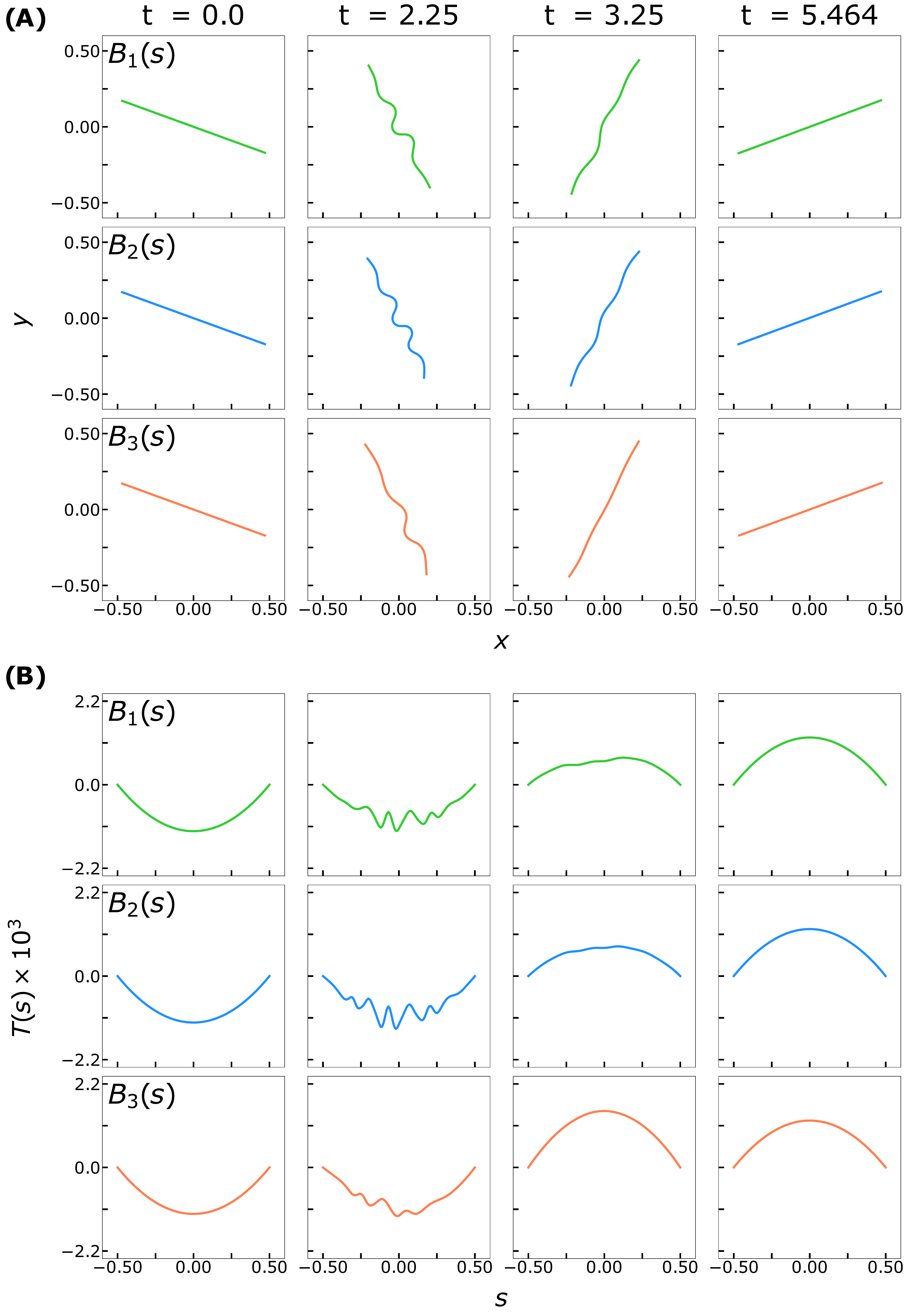}
 \caption{(A) Snapshots of filament buckling behavior in non-Brownian shear flow at various timepoints based on the three different rigidity profiles, all for flow strength $\mubar = 5 \times 10^{5}$. (B) The filament tension profile at the same timepoints as a function of the arclength $s$.}
 \label{fig:filament_shear_flow}
\end{figure}

We begin by describing our observations from non-Brownian simulations of a filament with the three different bending stiffness profiles shown in Figure~\ref{fig:3_rigidity_profiles}. We initially orient a filament along a straight line at an angle of $\theta = 8\pi/9$ relative to the horizontal and supply a perturbation of magnitude $\mathcal{O}(10^{-4})$ to the filament's $y$-component to induce buckling in shear flow $\Uvelocity = (y,0)$. In this configuration, the filament is placed in the compressive quadrants of shear flow, coinciding with a negative parabolic filament tension (Figure~\ref{fig:filament_shear_flow}B). As the filament rotates, the compressive forces acting upon the filament eventually overcome the internal elastic resistance to bending, causing the filament to buckle. During this process, the filament tension loses its parabolic profile. Once the filament is in the extensional quadrants, the internal filament tension is positive and stretches the filament: the rightmost panels show the configuration at $\theta = \pi/9$, approximately at $t = 5.464$. This behavior in shear flow has been well studied and documented for a constant stiffness profile. \cite{Tornberg2004} We observe qualitative differences between the filament configuration and tension by comparing the case of constant stiffness with the locally weak stiffness profile $B_{2}(s)$ and the asymmetrically rigid stiffness profile $B_{3}(s)$: as expected, the magnitude of tension fluctuations and deformations are larger for the locally weak profile. 

\begin{figure}[t]
 \centering
 \includegraphics[width=0.48\textwidth]{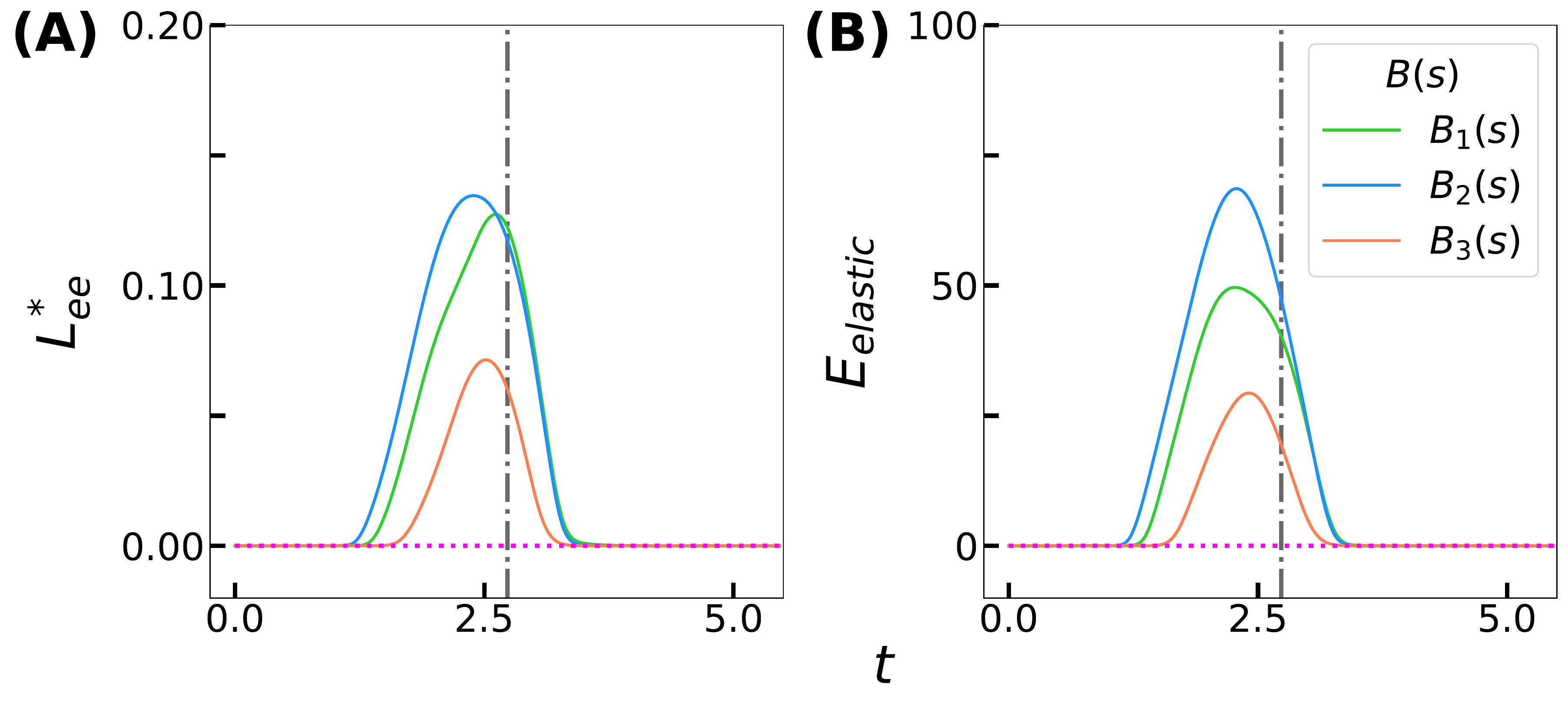}
 \caption{(A) Measured filament compression $L_{ee}^{*}$ and (B) elastic energy for the duration of the simulation of the three stiffness profiles in Figure~\ref{fig:filament_shear_flow}. The magenta dotted line in each panel represents a rigid, unperturbed filament. The gray dash-dot vertical line represents the time at which a rigid filament is oriented at $\theta = \pi/2$ relative to the horizontal.
 \label{fig:filament_simulation_energy_length}}
\end{figure}

\begin{figure}[ht]
 \centering
 \includegraphics[width=0.48\textwidth]{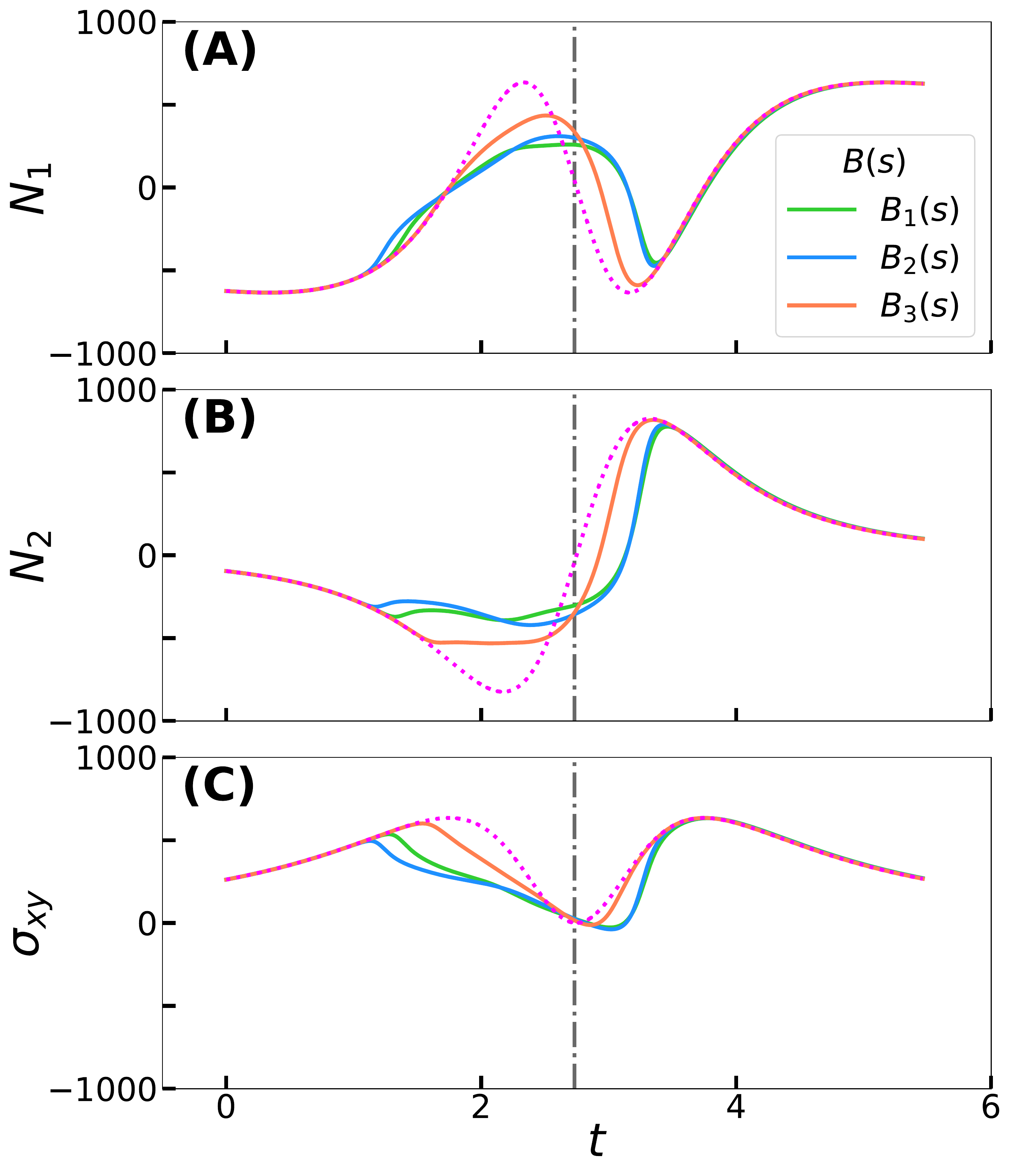}
 \caption{Calculated partcle extra stress contributions for the three different rigidity profiles during from Figure~\ref{fig:filament_shear_flow}: (A) first normal stress difference $N_{1}$, (B) second normal stress difference $N_{1}$, and (C) shear stress $\sigma_{xy}$. The magenta dotted line in each panel represents an unperturbed filament that rotates as a rigid rod. The gray dash-dot vertical line represents the time at which a rigid rod is oriented at $\theta = \pi/2$ relative to the horizontal.
 \label{fig:filament_simulation_stress}}
\end{figure}

These differences among the stiffness profiles can be better quantified by comparing the effective compression or the filament end-to-end length deficit in Figure~\ref{fig:filament_simulation_energy_length}A. This quantity is defined as $L_{ee}^{*} = 1-L_{ee}/L$, where $L_{ee}$ is the end-to-end distance. A locally weak filament is more compressed relative to a uniformly stiff filament, whereas the asymmetrically stiffer filament is more resistant to end-to-end compression. The same trends can be quantified by comparing the filament elastic energy ${E}_{elastic}= 1/2\int^{L/2}_{-L/2}{B(s)\xss^{2} \: ds}$. We observe that the filament compression trends are analogous to the trends in elastic energy in Figure~\ref{fig:filament_simulation_energy_length}B.

These deformations and storage or dissipation of elastic energy introduces rheological signatures in the suspended fluid. \cite{Batchelor_stess1970,Tornberg2004,Chakrabarti2021} To quantify the effects of these different stiffness profiles on the the stress system of the fluid containing the filament, we calculate the particle extra stress tensor \cite{Batchelor_stess1970}:
\begin{equation}\label{stress_tensor}
\boldsymbol{\sigma} = \displaystyle \frac{1}{2}\int^{L/2}_{-L/2} \left[ \force (s) \xloc (s) + \xloc (s) \force (s)\right]  \: ds.
\end{equation}
We show the evolution of the first and second normal stress differences, $N_{1} = \sigma_{xx} - \sigma_{yy}$ and $N_{2} = \sigma_{yy} - \sigma_{zz}$ respectively, in Figure~\ref{fig:filament_simulation_stress}A-B. The first normal stress difference is zero for a rigid rod rotating in shear flow and non-zero for buckled filaments over a full rotation.\cite{Tornberg2004,Becker2001,Chakrabarti2021} To confirm this, we quantify the total extra stress during a full deformation cycle by integrating the area under the curves from $t = 0$ to $t = 5.464$, such that the filament is approximately oriented at $\pi/2$ from the horizontal halfway through this period (Table~\ref{tab:stress_values}). We obtain a small but negative value for  $N_{1,tot}$ for a rigid rod (dotted pink line in Figure~\ref{fig:filament_simulation_stress}A). This could be attributed to our chosen initial filament orientation, choice of parameters, or absence of the non-local operator in Equation~(\ref{eq:filament_velocity_non_brownian_dimensionless}). Nonetheless, a buckled filament with a uniform stiffness profile yields a positive non-zero first normal stress difference, in agreement with previous studies.\cite{Tornberg2004,Becker2001,Chakrabarti2021} Interestingly, $N_{1,tot}$ for a locally weak $B_{2}(s)$ stiffness profile is larger than that for a uniformly stiff filament. In a confined flow geometry where the fluid is entrapped between walls, this corresponds to the fluid exerting more stress on the walls. $N_{1,tot}$ for a asymmetrically rigid $B_{3}(s)$ stiffness profile is also positive and non-zero, but smaller in magnitude than the other stiffness profiles. These results suggest that the extent of filament buckling is correlated with the magnitude of the first normal stress difference. Modifying the filament stiffness profile to favor buckling deformations will increase $N_{1,tot}$. The $N_{2,tot}$ trends are analogous to those of $N_{1,tot}$, where the total stress difference is the highest for a locally weak filament backbone.
\begin{table}
\caption{\label{tab:stress_values}Total value of stress types by integration of area under the curve in Figure~\ref{fig:filament_simulation_stress} for each stiffness profile. The unperturbed column represents a filament without a supplied perturbation so it rotates like a rigid rod in shear flow (corresponding to the dotted magenta lines in Figure~\ref{fig:filament_simulation_stress}}.
\begin{ruledtabular}
\begin{tabular}{cllll}
Stress Type & Unperturbed & $B_{1}(s)$ & $B_{2}(s)$ & $B_{3}(s)$ \\
\hline
$N_{1,tot}$ & -20.0 & 84.2 & 135.9 & 27.8\\
$N_{2,tot}$ & -3.1 & 12.6 & 14.7 & 1.90\\
$\sigma_{xy,tot}$ & 2276.9 & 1909.2 & 1859.5 & 2138.9\\
\end{tabular}
\end{ruledtabular}
\end{table} 

Plotting the evolution of shear stresses, $\sigma_{xy}$, over the duration of the simulation also reveals differences between the three stiffness profiles in Figure~\ref{fig:filament_simulation_stress}C. Tornberg \& Shelley\cite{Tornberg2004} reported that filament buckling reduces the shear stress in comparison to a rigid, unbuckled filament.  Excess local buckling, associated with a higher end-to-end length deficit and elastic energy in our simulations, could further reduce the shear stress perceived by the filament. To confirm this, we compute the total shear stress for each stiffness profile during the same time period as for the stress differences and report the values in Table~\ref{tab:stress_values}. With our chosen stiffness profiles, more buckling is correlated with lower shear stress.

\subsubsection{\label{sec:lin_stability_analysis}Linear Stability Analysis}

\begin{figure*}[ht]
 \centering
 \includegraphics[width=\textwidth]{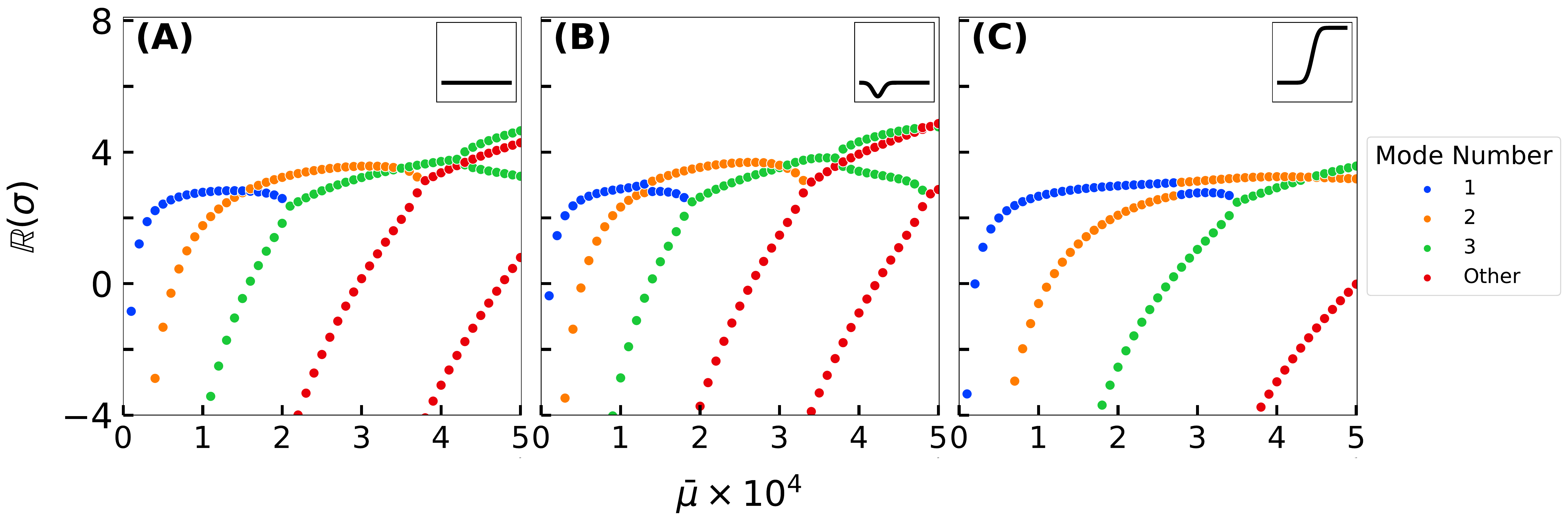}
 \caption{Eigenspectrum stability maps for three different rigidity profiles that highlight the stability of each mode number as a function of the elastoviscous number $\mubar$. Positive growth rates ($\mathbb{R}(\sigma) > 0$) correspond to unstable modes. Modes are color-coded based on their mode number with `Other Mode Number' indicating fourth and higher modes. Data plotted on the graphs are plotted in increments of 1000 $\mubar$. Insets show stiffness profiles corresponding to each eigenspectrum map.}
 \label{fig:linear_stability_eigenspectrum}
\end{figure*}

To help explain these observed differences from our simulations, we turn our attention to linear stability analysis for a heterogeneously stiff filament in extensional flow; we expect quantitative features from such an analysis to carry over to shear flow which is combination of extension and rotation. An unperturbed filament resting in along the x-axis of a 2D extensional flow profile of $\Uvelocity = (-x,y)$ adopts an unperturbed parabolic tension profile of the form $T(s) = \mubar(1/4-s^{2})/4c$.\cite{Kantsler2012,Batchelor1970} We define the unperturbed configuration of the filament as $\xloc(s,t) = (s,0)$. Acknowledging that $c \gg 1$ for slender fibers where $c = \ln(\epsilon^{2}e)$, we simplify Equation~(\ref{eq:filament_velocity_non_brownian_dimensionless}) as
\begin{equation}\label{eq:stability_centerline}
\mubar\left(\xt - \Uvelocity\right) = c\left(\identity + \xs\xs\right) \cdot \force.
\end{equation}
Perturbing Equation~(\ref{eq:stability_centerline}) with small vertical displacement $h(s,t)$ and neglecting higher order terms yields a linearized equation for perturbations: 
\begin{equation}\label{eq:stability_centerline_linearized}
\mubar\left(h_{t} - h\right) = c\left(-2T_{s}h_{s} - Th_{ss} + B_{ss}h_{ss} + 2B_{s}h_{sss} + h_{ssss}\right).
\end{equation}
In contrast with previous linear stability analyses, Equation~(\ref{eq:stability_centerline_linearized}) accounts for any modeled or fitted stiffness profile $B(s)$. We perform a normal mode analysis by setting $h(s,t) = \hhat(s)e^{\sigma t}$ where $\hhat(s)$ is the mode shape and $\sigma$ is the associated complex growth rate. Substituting the normal modes into Equation~\eqref{eq:stability_centerline_linearized} yields the an eigenvalue-eigenfunction (growth rate and mode shapes, respectively) problem:
\begin{equation}\label{eq:nonconstant_K_stability}
\begin{split}
B(s)\pdv[4]{\hhat}{s} + 2B_{s}\pdv[3]{\hhat}{s} + \left(B_{ss} - \frac{\mubar}{4c} \left[\frac{1}{4} - s^{2}\right]\right)\pdv[2]{\hhat}{s} \\
+ \frac{\mubar s}{c}\pdv[]{\hhat}{s} - \frac{\mubar}{c}(\sigma - 1)\hhat = 0.
\end{split}
\end{equation}
This problem is numerically solved by applying the torque-free and force-free boundary conditions of Equation~(\ref{eq:torque_force_boundary_condition_all}). Like our simulations, we approximate the derivatives with second-order finite differences. The eigenspectrum maps in Figure~\ref{fig:linear_stability_eigenspectrum} plot the real component of the growth rate $\sigma$ as a function of the flow strength $\mubar$ to describe the stability of mode shapes for each stiffness profile. We define the mode number of the filament shapes realized in our analysis as the number of inflections in the mode shape at the onset of stability and track them over a range of $\mubar$ values, but these patterns can drastically change, as we illustrate below.

In the case of a uniform stiffness profile, plotting the growth rate as a function of the flow strength reveals a familiar and well-studied landscape: \cite{Young2007,Guglielmini2012} increasing the flow strength sequentially destabilizes higher buckling modes. The first mode is typically `U' shaped and appears at the threshold $\mubar \approx 1258$ (Table~\ref{tab:buckling_threshold}). This onset of instability is consistent with previous findings.\cite{Young2007,Kantsler2012,Guglielmini2012,Chakrabarti2020} Increasing $\mubar$ to approximately $6358$ destabilizes the second mode, typically `S' shaped. The third mode, typically `W' shaped, becomes unstable at $\mubar \approx 15,850$. The stability curves for odd-numbered modes (first mode, third mode, etc.) have a tendency to merge with each other, physically corresponding to an additional `bump' appearing in the shape. For instance, the `U' shape develops a bump in the center around the merger with the `W' shape; likewise, even-numbered modes merge with each other (Figure \ref{fig:linear_stability_eigenspectrum}A). When these merger events happen, we label the resulting shape with the higher numbered mode.

\begin{table}
\caption{\label{tab:buckling_threshold} Critical flow strengths representing the onset of instability ($\sigma = 0$) for the first 3 modes for each rigidity profile.}
\begin{ruledtabular}
\begin{tabular}{clll}
Mode Number & $B_{1}(s)$ & $B_{2}(s)$ & $B_{3}(s)$ \\
\hline
1 & 1,258 & 1,112 & 2,005\\
2 & 6,358 & 5,135 & 11,245\\
3 & 15,851 & 13,740 &26,321\\
\end{tabular}
\end{ruledtabular}
\end{table} 

The non-uniform filament stiffness profiles lead to noticeable changes in their respective eigenspectrum maps. The locally weak $B_{2}(s)$ stiffness profile is characterized by a decrease in the rigidity on one side of the filament, potentially decreasing the overall stability of the filament. On the other hand, the asymetrically rigid $B_{3}(s)$ stiffness profile is characterized by a rapid increase in the rigidity on one side of the filament, potentially strengthening the filament against buckling. Such changes to the stiffness profiles result in an earlier or delayed onset of instability of the buckling modes as shown in Table~\ref{tab:buckling_threshold}, confirming our initial hypotheses about filament stability.

\begin{figure}[ht]
 \centering
 \includegraphics[width=0.48\textwidth]{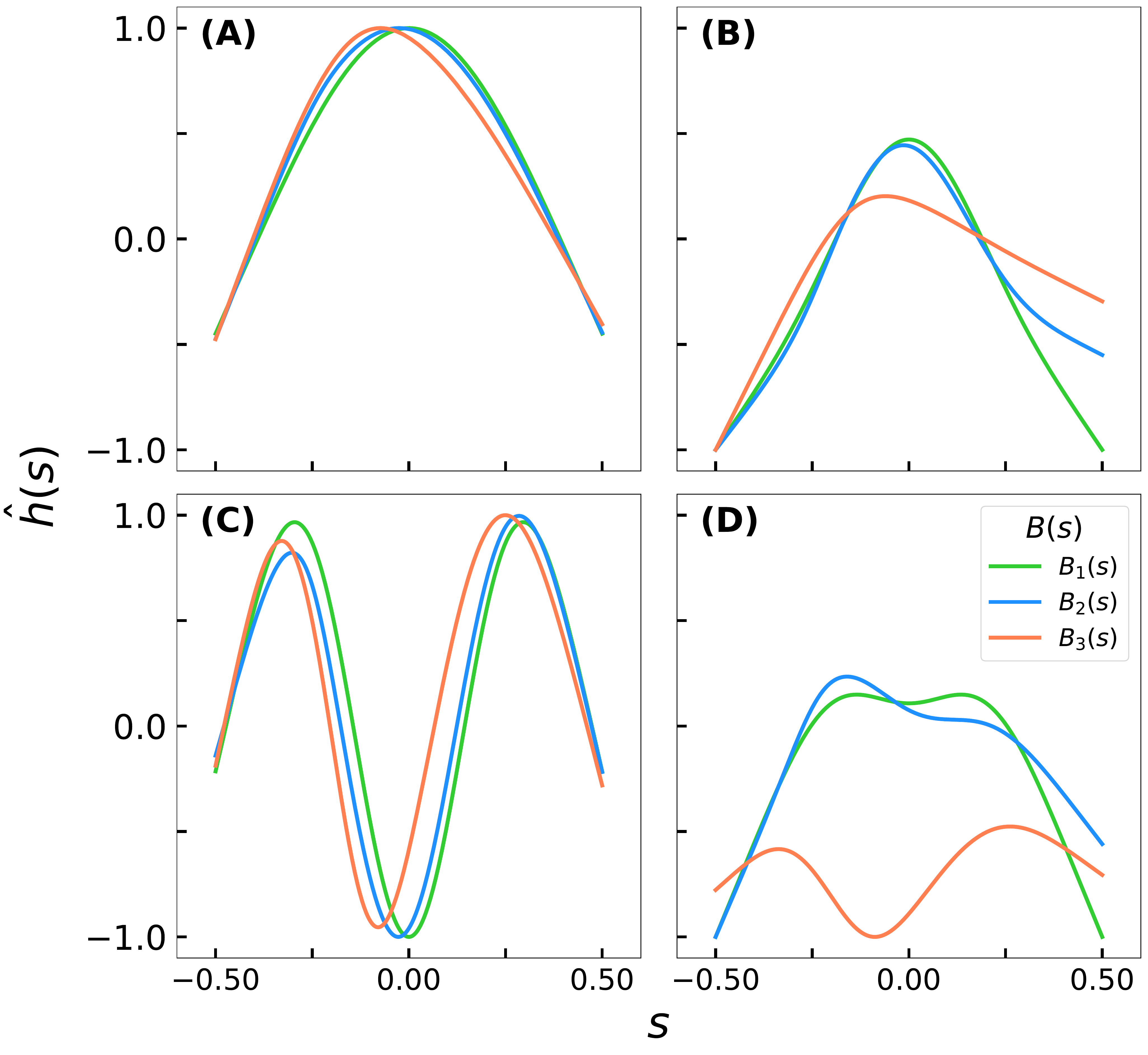}
 \caption{Comparison of the first and third mode shapes for each rigidity profile at various $\mubar$ values. A) and C): The first and third mode shapes at the onset of instability, respectively (see Table~\ref{tab:buckling_threshold} for critical $\mubar$ values). B) and D): The first and third mode shapes at flow strengths much larger than these critical values, at $\mubar = 18,000$ and at $\mubar = 30,000$ respectively. All mode shapes are normalized to a max amplitude of 1. }
 \label{fig:linear_stability_different_shapes}
\end{figure}

To examine the evolution of the predicted mode shapes between the different stiffness profiles, we compare $\hhat(s)$ for each stiffness profile at various $\mubar$ values. Examining the first mode shape at its onset of instability reveals minute differences in the shape between the different stiffness profiles (Figure~\ref{fig:linear_stability_different_shapes}A). However, we start to see differences in the shape of the first mode and visible effects of asymmetry upon increasing the flow strength significantly beyond the critical $\mubar$ corresponding to that mode (Figure~\ref{fig:linear_stability_different_shapes}B). Similarly, comparing the third mode (Figure~\ref{fig:linear_stability_different_shapes}C-D), we see that the difference in stiffness profiles results in minute differences near critical thresholds, whereas ramping up the flow strength reveals dramatic changes and asymmetries in the mode shape. 

For all the discussion so far, we must note that linear stability analysis predicts the most dominant mode shapes and mode numbers at specific $\mubar$ values in a deterministic simulation. The non-linear shapes seen in the simulations, however, depend on the perturbations applied to the filament. In particular, real microfilaments are subject to Brownian fluctuations and so we can at best make statistically expected predictions of buckling shapes or thresholds. Indeed, Brownian fluctuations are a source of constant perturbations and excite all modes equally,\cite{Kantsler2012} although excited modes are expected to be consistent with the deterministic predictions.\cite{Manikantan2015} In what follows, we develop a framework to address stochastic buckling of Brownian microfilaments and to quantify the role of thermal fluctuations and non-uniform rigidity on the statistically expected stability and shape.

\subsubsection{\label{sec:simulation_stability_comparison}Comparison of Simulations with Stability Analysis}

\begin{figure*}[ht]
 \centering
 \includegraphics[width=0.98\textwidth]{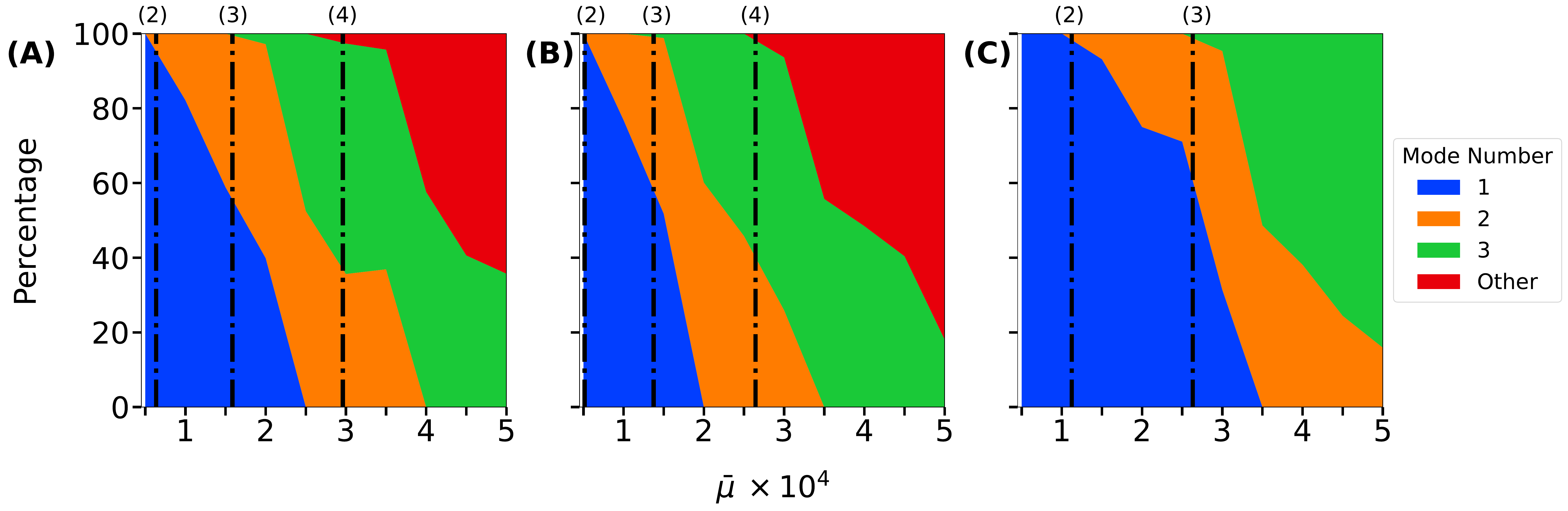}
 \caption{Distribution of the most unstable mode in extensional Brownian simulations in instances where the filament crossed the Brownian noise floor threshold for (A) $B_{1}(s)$, (B) $B_{2}(s)$, and (C) $B_{3}(s)$. The dot-dashed lines on each graph, denoted with $(2)$, $(3)$, or $(4)$ represent the calculated onset of instability for the second, third, and fourth modes respectively (see Table~\ref{tab:buckling_threshold} for the critical $\mubar$ values).}
 \label{fig:most_excited_modes}
\end{figure*}

We will compare the predictions of the stability analyses with filament shapes that emerge in nonlinear and Brownian simulations at short times. As an illustration, we perform Brownian ensemble simulations (statistical averages across $200$ simulations for each data point) for a moderately rigid filament ($\ell_{p}/L = 100$) with the three different stiffness profiles in extensional flow. All modes are excited by thermal noise, and thus one way to extract growth rates is by projecting the deflections obtained in numerical simulations on a complete basis of orthogonal shape functions. However, the linear operator in Equation~(\ref{eq:nonconstant_K_stability}) is not self-adjoint and the eigenfunctions associated with the normal mode analysis do not form an orthogonal basis. \cite{Guglielmini2012,Chakrabarti2020} Equation~(\ref{eq:nonconstant_K_stability}) can be written as a classic eigenvalue-eigenfunction problem
\begin{equation}\label{eq:stability_eigen_formulation}
\frac{\mubar}{c} \lambda \phi = \loperator[\phi],
\end{equation}
where the linear operator is
\begin{equation}\label{eq:linear_operator_non_constant_K}
\begin{aligned}	
\loperator[\phi] = B(s)\pdv[4]{\phi}{s} + & 2B_{s}\pdv[3]{\phi}{s} + \left(B_{ss} - \frac{\mubar}{4c} \left[\frac{1}{4} - s^{2}\right]\right)\pdv[2]{\phi}{s}  \\
& + \frac{\mubar s}{c}\pdv{\phi}{s} + \frac{\mubar}{c}\phi
\end{aligned}	
\end{equation}
and $\loperator$ satisfies the boundary conditions for the linear stability problem (Section~\ref{sec:lin_stability_analysis}). Then, the family of eigenfunctions $\phi$ do not form an orthogonal basis for the shapes that the filament can take. Here, we use $\phi$ to denote the eigenfunctions of the linear operator and are identical to the mode shapes $\hhat$ of linear stability analysis. We denote $\lambda$ as the eigenvalues associated with this problem, which are identical to the growth rate $\sigma$ from linear stability analysis. 

Thus, we seek the adjoint $\loperator^{\dag}$ of the linear operator $\loperator$. Borrowing notation from Chakrabarti \textit{et al.},\cite{Chakrabarti2020} the adjoint operator $\loperator^{\dag}$ is defined as
\begin{subequations}\label{eq:adjoint_definition}
\begin{equation}\label{eq:linear_adjoint_definition}
\langle v,\loperator [w] \rangle = \langle\loperator^{\dag} [v], w\rangle,
\end{equation}
\begin{equation}\label{eq:linear_adjoint_bracket_term_definition}
\langle v,\loperator [w] \rangle = \displaystyle \int^{+1/2}_{-1/2}{v \loperator [w]\: ds}.
\end{equation}
\end{subequations}
Here, $w$ represents eigenfunctions that satisfy the boundary conditions for the linear stability problem (Section~\ref{sec:lin_stability_analysis}), and $v$ represents eigenfunctions that are adjoint to $w$. Repeated integration by parts to satisfy Equation~(\ref{eq:linear_adjoint_definition}) reveals the following adjoint operator:
\begin{equation}
\begin{aligned}
\loperator^{\dag}[\Phi] = & B(s)\pdv[4]{\Phi}{s} + 2B_{s}\pdv[3]{\Phi}{s} + \left(B_{ss} - \frac{\mubar}{4c}\left[\frac{1}{4} - s^{2}\right]\right)\pdv[2]{\Phi}{s} \\
& + \frac{\mubar}{2c}\Phi, \label{eq:adjoint_operator_non_constant_K}
\end{aligned}
\end{equation}
with corresponding boundary conditions:
\begin{subequations}\label{eq:adjoint_boundary_conditions_all}
\begin{equation}\label{eq:adjoint_boundary_conditions_lhs}
s = -1/2: B(s) \Phi_{ss} = -B(s) \Phi_{sss} - \frac{\mubar}{4c}\Phi = 0,
\end{equation}
\begin{equation}\label{eq:adjoint_boundary_conditions_rhs}	
s = +1/2: B(s) \Phi_{ss} = -B(s) \Phi_{sss} + \frac{\mubar}{4c}\Phi = 0,
\end{equation}	
\end{subequations}
where $\Phi$ is the adjoint eigenfunction. Throughout this formulation, both $\phi$ and $\Phi$ share the same eigenvalues. And, by definition, eigenfunction pairs $\phi_{i}$ and $\Phi_{i}$ are orthogonal; we can calculate the normalization constant $C_{i}$ between the two eigenfuctions by defining an inner product:
\begin{equation}\label{eq:normalization_constant_definition}
\int^{1/2}_{-1/2}{\phi_{i} \Phi_{j} ds} = C_{i}\delta_{ij},
\end{equation}
where $\delta_{ij}$ is the Kronecker delta. 

Between the family of eigenfunctions $\phi$ and the orthogonality condition in Equation~(\ref{eq:normalization_constant_definition}), we now have a complete basis and orthogonality condition on which we can project any perturbed filament shape. We do this by writing the shape of the filament $\hhatsim$ obtained during extensional flow from non-linear simulations as a linear combination of all the eigenfunctions weighted by an amplitude $a_{i}$:
\begin{equation}\label{eq:simulation_shape_equation}
\hhatsim = \sum^{\infty}_{i = 1}{a_{i}(t)\phi_{i}(s)}.
\end{equation}	
The amplitude corresponding to each eigenfunction can be extracted using the orthogonality condition of Equation~(\ref{eq:normalization_constant_definition}):
\begin{equation}\label{eq:solve_for_amplitude}
a_{i} = \frac{1}{C_{i}}\int^{L/2}_{-L/2}{\hhatsim \Phi_{i} \:ds}.
\end{equation}
In the linear short-time regime, these amplitudes grow as
\begin{equation}\label{eq:amplitude_growth_rate}
a_{i}(t) = A_{i}e^{\sigma_{i} t},
\end{equation}
where $A_{i}$ is a constant. This formulation allows us to: 1) formally quantify the most unstable mode in Brownian nonlinear simulations based on the extracted $\sigma_{i}$ values, and 2) compare these growth rates from simulations with the linearized and deterministic predictions of our stability analysis.

\begin{figure*}[t!]
 \centering
 \includegraphics[width=0.98\textwidth]{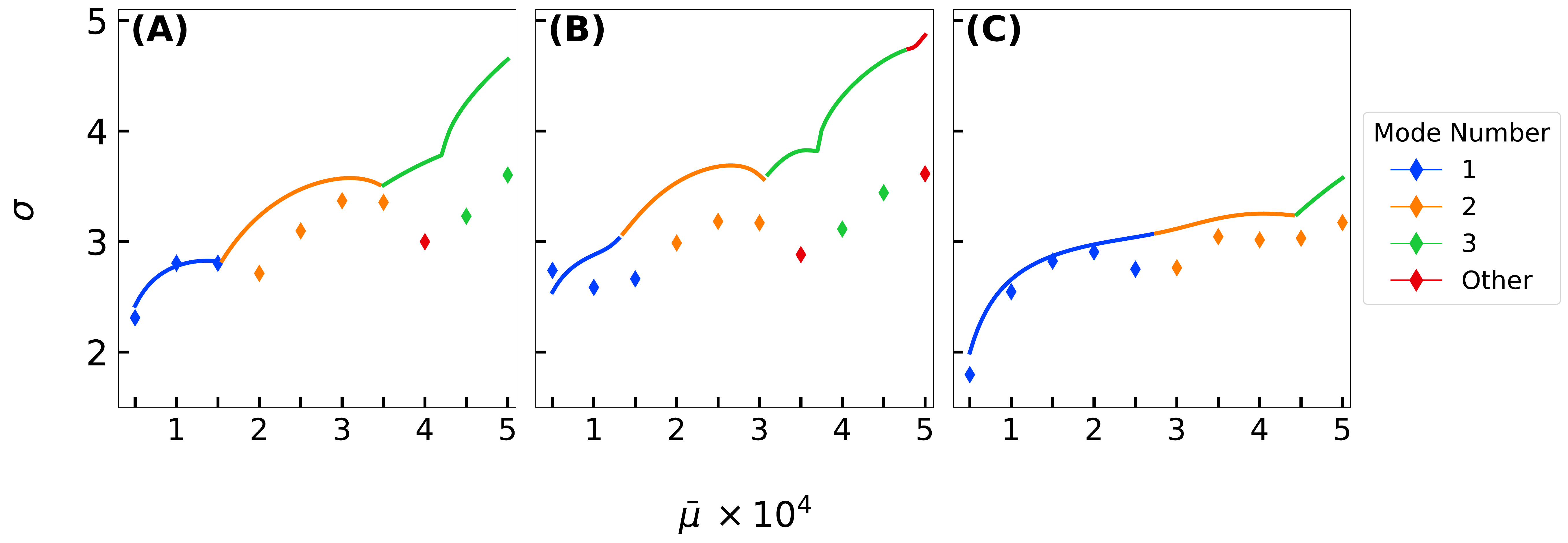}
 \caption{Comparison of the predicted growth rates obtained from linear stability analysis against the calculated growth rates from the Brownian extensional simulations at $\ell_{p}/L=100$ along with their corresponding mode number classifications for (A) $B_{1}(s)$, (B) $B_{2}(s)$, and (C) $B_{3}(s)$. The highest growth rates from either stability analysis or simulations are plotted. Lines represent the stability analysis predictions. Diamond points represent the extracted growth rate from the simulations.\label{fig:simulation_stability_comparisons}}
\end{figure*}

Before such a comparison can be made, we must note that Brownian fluctuations excite all of these orthogonal modes.\cite{Kantsler2012,manikantan_phd2015} The buckling amplitudes due to flow can only be reasonably extracted above a Brownian noise floor. We thus need a statistical estimate of the filament's amplitude due to Brownian fluctuations alone. We can approximate this noise floor as the expected value of Brownian fluctuations in an extensional flow. We follow Kantsler \& Goldstein's\cite{Kantsler2012}  notations to represent the elastic and tension energy of a filament due to small fluctuations $h(x)$ as:
\begin{equation}\label{eq:energy_noise_floor}
\mathcal{E} = \frac{1}{2} \displaystyle \int^{L/2}_{-L/2}{(\kappa(s) h_{xx}^{2} + T(s)h_{x}^{2}) \: ds},
\end{equation}
where the tension is of the form:
\begin{equation}\label{eq:energy_noise_floor_tension}
T (s) = \frac{2\pi\mu \dot{\gamma}}{\ln(1/\epsilon^{2}e)}\left[\frac{L^{2}}{4}-x^{2}\right].
\end{equation}
In Equation~\eqref{eq:energy_noise_floor}, we assume the rigidity of the filament $\kappa (s)$ to be constant and uniform to estimate a baseline noise floor. Integrating Equation~(\ref{eq:energy_noise_floor}) by parts repeatedly using the force-free and torque-free boundary conditions and projecting $h(x)$ onto the orthogonal eigenfunction $\phi_i(x)$ basis as $h(x) = \sum^{\infty}_{n=1}a_{n}\phi_{n}(x)$ with eigenvalues $\lambda_{n}$ gives\cite{manikantan_phd2015}
\begin{equation}\label{eq:energy_sum}
\mathcal{E} = \frac{L}{2} \displaystyle \displaystyle \sum_{n=1}^{\infty}{a_{n}^{2}\lambda_{n}}.
\end{equation}
From the equipartition principle, each independent mode in the summation then contributes $k_{B}T/2$ towards the total energy. Rescaling the eigenvalues to $\Lambda_{n} = \lambda L^{4}/\pi^{4}\kappa$ and using the persistence length $\ell_{p} = \kappa/k_{B}T$ gives:
\begin{equation}\label{eq:energy_contribution_modes}
\langle a_{m}a_{n} \rangle  = \delta_{mn}\frac{L^{2}}{\Lambda_{n}\pi^{4}}\frac{L}{\ell_{p}}
\end{equation}
The values of $\Lambda_{n}$ are approximated as $(n+0.5)^{4}$ from numerical calculations,\cite{Wiggins1998,Kantsler2012,manikantan_phd2015} providing a statistical estimate for the amplitude due to thermal fluctuations alone. We filter for extracted amplitudes $a_{i}$ above the first mode's noise floor.

Next, we individually fit each ensemble's amplitude above this noise floor to an exponential in time. The linear stability analysis and corresponding growth rates use the flow rate $\dot{\gamma}$ as the characteristic time scale, whereas the Brownian simulations use the much smaller filament relaxation timescale. To accommodate this change, the appropriate exponential fit for the amplitude from simulations that match the linear stability predictions must be of the form $a_i\sim \exp(\sigma_{i} \tau)$, where $\tau=\mubar t^{Br}$ and $t^{Br}$ is the dimensionless Brownian time scale in simulations. We fit $\ln(\lvert a_{i} \rvert)$ versus $\tau$ and select for modes whose amplitudes linearly grow with time ($R^{2} \geq 0.60$ where $R^{2}$ is the coefficient of determination from the linear fits) and calculate their respective growth rates. This fit is only done for short times, as defined by $\tau \leq 0.75$. The distribution of the fastest growing mode across a spectrum of flow strengths for each bending stiffness profile is summarized in Figure~\ref{fig:most_excited_modes}. The most dominant shapes observed in our simulations are consistent with the deterministic predictions from stability analysis, with the lower modes becoming less dominant at higher flow strengths in favor of higher modes. Additionally, we see effects of the rounded transitions of the shapes due to thermal fluctuations, consistent with previous findings. \cite{Baczynski2008,Manikantan2015}  Note, however, the effects of the rounded mode transitions and evolution of most modes are limited by the number of ensembles (200 simulations each) and resolution of evaluated $\mubar$ values (shown in increments of $5000$).

To quantitatively compare stability analysis with simulations, we average out our ensemble data by computing the root-mean-squared amplitude across ensembles and extracting a growth rate like above. Like our previous analysis, we filter for amplitude values above the noise floor and on short time scales. Figure~\ref{fig:simulation_stability_comparisons} shows the growth rate of the most unstable mode from the average ensemble simulation data plotted along with the maximum growth rate from our linear stability analyses. Again, we see that stability predictions and stochastic nonlinear simulations are fairly consistent in the mode that is most excited as well as in the magnitude of the growth rate.

\section{\label{sec:level1}Conclusions}

In summary, we have developed a framework for slender-body theory to incorporate non-uniform bending stiffness of flexible filaments. We use this platform to examine the flow-induced buckling behavior of realistic microfilaments with a non-uniform elastic backbone. As a demonstration of the method and mathematical tools, we model the adsorption or desorption of proteins to result in a locally or asymmetrically weaker or stiffer microfilament. We study the effects of the altered stiffness profiles in simple shear flow, where regions of modified filament backbone give rise to differences in the buckling patterns, tension, elastic energy, and stress tensor components. By comparing the short-time evolution of our Brownian and nonlinear simulations to linear stability analyses for each of the $\kappa(s)$ profiles, we are able to highlight features of enhanced or reduced local filament stiffness. 

The model stiffness profiles considered in this work allowed us to arrive at tractable filament shapes and illustrate a consistent set of steps to extract mode shapes from simulations. However, we emphasize that the mathematical machinery developed here is applicable to any modeled or experimentally measured profile of the filament rigidity. We have established the fundamental basis for this direction of analysis of stiff biopolymers such as microtubules with non-uniform protein condensation. \cite{Hernandez_Vega2017,Setru2021}  It remains to be seen how this description carries over to more complex flows or to substrate attachments that arise in the experimental tracking of such microtubules \cite{Tan2019}, and how these differences in buckling behavior translate to large-scale filament dispersion \cite{Manikantan2013} or cross-streamline migration. \cite{Xue2022} Our model can be readily extended to temporal variations in filament stiffness as well, setting the stage for future work on finite-time kinetics of protein adsorption/desorption and the elastohydrodynamics associated with such a process. We anticipate that the tools and insights gathered in the current work will support these future research directions.

\section{Acknowledgements}
We acknowledge financial support from a GAANN fellowship (T.N.) and a Hellman Fellowship (H.M.).

\section*{References}
\vspace{-4mm}

\bibliography{sbt_K_ref}

\end{document}